\documentclass[sn-aps,Numbered]{sn-jnl}

\usepackage{graphicx}%
\usepackage{multirow}%
\usepackage{amsmath,amssymb,amsfonts}%
\usepackage{amsthm}%
\usepackage{mathrsfs}%
\usepackage[title]{appendix}%
\usepackage{xcolor}%
\usepackage{textcomp}%
\usepackage{manyfoot}%
\usepackage{booktabs}%
\usepackage{algorithm}%
\usepackage{algorithmicx}%
\usepackage{algpseudocode}%
\usepackage{listings}%
\usepackage{latexsym}
\usepackage{graphicx}
\usepackage{dcolumn}
\usepackage{bm}
\usepackage{color}
\usepackage{titlesec}
\usepackage{float}
\usepackage{units}

\renewcommand{\vec}[1]{\mathbf{#1}}

\newcommand{\vp}{\vec v^{\mathrm{p}}}
\newcommand{\vs}{\vec v^{\mathrm{s}}}

\newcommand{\hp}{\eta^{\mathrm{p}}}
\newcommand{\hs}{\eta^{\mathrm{s}}}

\usepackage{stackengine}
\stackMath

\begin{document}

\title[Model chromatin flows: numerical analysis of linear and nonlinear hydrodynamics inside a sphere]{Model chromatin flows: numerical analysis of linear and nonlinear hydrodynamics inside a sphere}
\author[1]{\fnm{Iraj} \sur{Eshghi}}\email{ie355@nyu.edu}

\author[1]{\fnm{Alexandra} \sur{Zidovska}}\email{az45@nyu.edu}

\author[1]{\fnm{Alexander Y.} \sur{Grosberg}}\email{ayg1@nyu.edu}

\affil[1]{\orgdiv{Center for Soft Matter Research, Department of Physics}, \orgname{New York University}, \orgaddress{\city{New York},  \state{NY},\postcode{10003}, \country{USA}}}

\date{\today}
\abstract{We solve a hydrodynamic model of active chromatin dynamics, within a confined geometry simulating the cell nucleus. Using both analytical and numerical methods, we describe the behavior of the chromatin polymer driven by the activity of motors having polar symmetry, both in the linear response regime as well as in the long-term, fully nonlinear regime of the flows. The introduction of a boundary induces a particular geometry in the flows of chromatin, which we describe using vector spherical harmonics, a tool which greatly simplifies both our analytical and numerical approaches. We find that the long-term behavior of this model in confinement is dominated by steady, transverse flows of chromatin which circulate around the spherical domain. These circulating flows are found to be robust to perturbations, and their characteristic size is set by the size of the domain. This gives us further insight into active chromatin dynamics in the cell nucleus, and provides a foundation for development of further, more complex models of active chromatin dynamics.}
\maketitle
\section{Introduction}
The cell nucleus houses the genome, containing genetic information needed for the cell's life \cite{alberts2017molecular}. This information is encoded into a long DNA molecule, which in cells forms a complex with histone proteins, the chromatin fiber \cite{van2012chromatin}. The cell nucleus encompasses chromatin as well as a variety of molecules such as proteins and RNA \cite{alberts2017molecular,spector1993macromolecular}, and is maintained out-of-equilibrium by a large number of active processes e.g. transcription, replication, chromatin remodeling and DNA repair \cite{rippe2007dynamic,zidovska2020rich}. Active processes affect both the dynamics \cite{hubner2010chromatin,fierz2019biophysics,zidovska2020self} and organization of the genome \cite{hubner2013chromatin,bickmore2013genome,mirny2022mechanisms}. Examples include changes in: chromatin mobility with transcription \cite{soutoglou2007mobility,lanctot2007dynamic}, local chromatin packing and dynamics upon DNA damage \cite{,mine2013dna,lebeaupin2015chromatin,eaton2020structural}, occasional directed motion of chromosomal loci \cite{marshall1997interphase,levi2005chromatin,chuang2006long}, or the active extrusion of chromatin loops by cohesin \cite{mirny2021keeping,davidson2021genome}. In particular, active processes were shown to lead to the formation of micron-scale chromatin domains, within which chromatin moves coherently \cite{zidovska2013micron}.

The first theoretical model developed to explore the origins of the chromatin coherent motions was based on two-fluid hydrodynamics \cite{bruinsma2014chromatin}, it was followed by explicit hydrodynamic computational models \cite{saintillan2018extensile,Mahajan2022hetero}. These papers led to the important insight that local active forces mediated through long-range hydrodynamic interactions can lead to large-scale motions of chromatin. Separately, several computational models were developed which do not include hydrodynamic interactions, but reproduce the phenomenology of coherent motions at the expense of introducing artificial long-range forces between sections of the polymer \cite{shi2018interphase,liu2018chain,di2018anomalous}.

An analytical description of chromatin hydrodynamics was developed in the works \cite{bruinsma2014chromatin,eshghi2022symmetry,eshghi2023alignment} based on the two-fluid model \cite{doi1992dynamic} and the ideas of statistical physics of active systems \cite{marchetti2013hydrodynamics,vicsek1995novel,toner1995long,AdarJoanny2021}. The most significant flows were obtained under the assumption that active motors driving are such that each of them exerts one force on the polymer and equal but opposite force on the solvent. Indeed typical active motors operating on the genome can be viewed as such, for instance, RNA polymerase II which is a common active enzyme in the nucleus \cite{kimura1999quantitation}. Therefore, in both our previous work \cite{eshghi2023alignment} as well as in the present paper, we consider our theory in the most general phenomenological form, not specifying the nature of motors beyond the fact that they have a polar symmetry. It is because of polar symmetry that they act on the relative velocity of the solvent past polymer like force monopoles and not dipoles, thus generating very significant flows. If the number and activity of motors exceeds a critical threshold, we found that they spontaneously form an ordered polar phase which actively pumps chromatin and solvent through one another. We successfully described this spontaneous ordering through a polar order parameter, and identified the value of this critical threshold as a function of model parameters along with the critical exponents near the transition \cite{eshghi2023alignment}.

These results, however, were found with the assumption of an infinite boundless medium. While this made the analysis more straightforward, a more accurate description of the situation experienced by the chromatin polymer would account for the finite size of the cell nucleus. In this work, we will study the dynamics of our model in a confined spherical geometry, near its critical point. We show that the critical value of force and density of motors is shifted by an amount dependent on model parameters and system size. Furthermore, we find that the length scale of the modes excited near the transition point is set by the confinement size, consistently with expectations from the equilibrium theory of second-order phase transitions.

\section{The model, confining geometry, and equations of motion}

Our goal in the present work is to analyze previously derived equations of hydrodynamic motion of chromatin \cite{eshghi2023alignment}, both linearized and nonlinear, when confined in a spherical domain. To make this work self-contained, and to set up the notations, we start by re-stating the primary equations of motion of this model. 

\subsection{Equations of motion}
We consider two mutually permeating fluids: a polymer and a solvent. Their velocities are denoted by the fields $\vp,\; \vs$ respectively, although in our calculations we will use the following two linear combinations of the velocity fields: $\vec w = \vp - \vs,\; \vec u = (\hp \vp + \hs\vs)/(\hp + \hs)$. The two fluids, flowing past another, experience a friction per unit volume $\zeta$. The volume fraction of the polymer is $\phi(\vec r)$, and the two-fluid combination is assumed to be incompressible, so the volume fraction of solvent is $1-\phi(\vec r)$. Both fluids are assumed to experience viscous dissipation upon shear, with respective viscosities $\hp,\;\hs$. The viscosity of the chromatin polymer is known to have a frequency dependence \cite{eshghi2021interphase}, but here we will assume that it is simply Newtonian. The polymer experiences an osmotic pressure $\Pi(\phi)$, which is only a function of $\phi$ as it is assumed to equilibrate quickly and locally. We consider a regime where the polymer density deviates weakly from its mean value $\phi_0$, allowing us to linearize the equations of motion around that point: $\phi = \phi_0 + \delta\phi$. As a result the osmotic pressure can be written as $\Pi = \Pi_0 + K\delta\phi$, where $K$ is the osmotic modulus of the polymer. This gives us the following equations for the fluids
\begin{subequations}
\begin{align}
\begin{split}
   & \zeta\left(\frac{1}{\hp}+\frac{1}{\hs}\right) \vec w   =  \nabla^2\vec w + \left( \frac{\vec F^{\mathrm{p}}}{\hp} - \frac{\vec F^{\mathrm{s}}}{\hs} \right)\\
    & \ \ \ \ \ \ \ \ \ \ \ +  \left(\frac{1-\phi_0}{\hs} - \frac{\phi_0}{\hp} \right) \nabla P - \frac{K}{\hp}\nabla\delta\phi \ ,
    \label{eq:eqmot_w_full}
\end{split}\\
    & \nabla \cdot \vec w = -\partial_t\delta\phi\left(\frac{1}{\phi_0} + \frac{1}{1-\phi_0}\right) \ ,
    \label{eq:cont_w}\\
    & \left( \hp + \hs \right) \nabla^2\vec u =    K \nabla\delta \phi + \nabla P + \vec F^{\mathrm{p}} + \vec F^{\mathrm{s}} \  ,
    \label{eq:eqmot_u_full}\\
    & \left( \hp + \hs \right)  \nabla \cdot \vec u = -\partial_t\delta\phi\left(\frac{\hp}{\phi_0} - \frac{\hs}{1-\phi_0}\right) \ ,
    \label{eq:cont_u}
    \end{align}
    \label{eq:fluid_eqmots_full}\\
\end{subequations}
where $\vec F^{\mathrm{p}},\;\vec F^{\mathrm{s}}$ are the forces densities generated by the motors, and $P$ is the hydrostatic pressure induced by total incompressibility.

As stated in the introduction, we assume that every active motor exerts a force on the polymer and equal and opposite force on the solvent. In the coarse-grained description, assuming there are some $\rho$ active motors per unit volume, we can write 
\begin{equation} 
    \vec F^{p} (r) = - \vec F^{s}(r) = \rho f \vec m(r) \ , 
\end{equation} 
where $f$ is the force produced by a single motor, while $\vec m(r)$ is the average vector of orientation of motors located around point $\vec r$. According to equations (\ref{eq:eqmot_w_full}) and (\ref{eq:eqmot_u_full}), field $\vec w$ is driven by force monopole density $f \rho \vec m$, while field $\vec u$ in this approximation is not driven at all.

Vector field $\vec m(\vec r)$ plays an important role in our theory, and it is worth several comments. First, the very possibility to define vector $\vec m$ is because motors we consider have polar symmetry, they have two different sides, one attached to the polymer and another exposed to the solvent. Second, vector $\vec m$, as the average of the unit orientation vectors of individual motors, has magnitude that is always smaller than unity: $|\vec m| \leq 1$. Third, this vector naturally serves as an order parameter of polarization ordering phase transition predicted and described by our theory.

\subsection{Linear Dynamics}

Given the polar symmetry of motors, each of them is subject to a torque whenever there is a flow of the solvent relative to the polymer, i.e., when $\vec w \neq 0$.  If drive is weak (because motors are either weak or not numerous enough), then this torque is also weak and the distribution of motor orientations is nearly isotropic.  In this case, the dynamics of average orientation is described by
\begin{equation}
    \tau \partial_t \vec m = -2\vec m + \frac{2\tau}{3a}\vec w\;,
    \label{eq:linear_m_eq}
\end{equation}
where we have introduced the reorientation time of the dipoles $\tau = \gamma/T$. $\gamma$ is the rotational friction coefficient of the motor, $T$ is the ambient (potentially effective) temperature, and $a$ is the size of the motors, assumed to be comparable to (or smaller than) the mesh size of the polymer. We can now combine equations (\ref{eq:linear_m_eq}) and (\ref{eq:fluid_eqmots_full}) into one equation for $\vec w$ and $\vec m$. 
\begin{equation}
\begin{split}
    &\left(1 + \lambda^2\nabla\times\nabla\times - \lambda_s^2\nabla\nabla\cdot\right)\tau\partial_t\vec w \\
    &-2\lambda_d^2\nabla\nabla\cdot \vec w_{\omega} = 
    \frac{f\rho\tau}{\zeta}\partial_t\vec m_{\omega}\;,
    \label{eq:general_eqmot_w}
\end{split}
\end{equation}
where we have defined three length scales:
\begin{equation} \lambda^2 = \frac{\hs}{\zeta},\quad\lambda_s^2 \simeq \frac{  \hp \left(1- \phi_0 \right)^2 }{\zeta },  \ \ \text{and} \ \ \lambda_d^2 \simeq \frac{K \phi_0 \left( 1 - \phi_0 \right)^2\tau}{2 \zeta} \ .
\end{equation}
The first of these, $\lambda$, is naturally identified as the mesh size of the polymer. Second, significantly larger length scale is $\lambda_s$, the screening length of hydrodynamic interactions in the two-fluid system. Finally, the third length scale, $\lambda_d$, characterizes the interplay of the two fluid system with motors, namely, it is the typical distance of cooperative diffusion by the polymer during motor reorientation time $\tau$; of course, this motion can be thought of as driven by osmotic elasticity of the polymer (described by $K$) and opposed by friction (described by $\zeta$). Since we are treating chromatin in a continuum approximation, the model we are working with is only relevant on length scales larger than the mesh size $\lambda$.

Since all of the equations of motion in this regime are linear, we can simultaneously decompose the fields $\vec u,\;\vec w,\;\vec m$ into their divergence-free (transverse, $\perp$) and curl-free (longitudinal, $\parallel$) components. Separating these into their respective equations of motion, we obtain
\begin{subequations}
\begin{align}
    \qquad \quad \left(1-\lambda_s^2 \nabla^2\right) \vec w_{\perp} &= \frac{f\rho}{\zeta}\vec m_{\perp}
    \label{eq:w_m_transverse}\\
    \label{eq:w_m_longitudinal}
    \left(1 - \lambda_s^2\nabla^2\right)\tau\partial_t\vec w_{\parallel} &= 2\lambda_d^2\nabla^2 \vec w_{\parallel} + \frac{f\rho\tau}{\zeta}\partial_t \vec m_{\parallel}\;.
\end{align}
\label{eq:eqmot_split}\\
\end{subequations}
Our previous study allowed us to identify the value of the force $f$ above which an ordered phase for $\vec m$ spontaneously develops. In an infinite domain, this occurs at $ f = 3a\zeta/\rho\tau$, where the velocity generated by the force dipoles $f\rho/\zeta$ equals the velocity needed to orient them $a/\tau$. This allows us to identify the critical parameter 
\begin{equation}
    \epsilon = \frac{f\rho\tau}{3a\zeta}-1\;.
\end{equation}
When $\epsilon > 0$, the zero modes of the velocity and orientation fields become unstable. This all changes, however, with the addition of boundaries, which set a maximal length scale for any dynamics for the two fluids. Our goal in this paper is to refine the conclusions of our previous work in the finite domain context.
\subsection{Nonlinear Dynamics}
Once an instability develops, the polar order parameter grows in magnitude exponentially. Eventually, nonlinear effects inevitably kick in. In particular, they guarantee that $|\vec m| \leq 1$. The precise evolution of $\vec m$ in the nonlinear regime is complex, as it couples to all the other moments of the distribution of motor orientations. However, in our previous work \cite{eshghi2023alignment} we derived an approximate equation of motion for $\vec m$ including only a second-order nonlinearity in $\vec w$:
\begin{equation}
\begin{split}
	\tau\partial_t \vec m &= 2(\vec m_{eq}(\vec w)- \vec m)\\
	\vec m_{eq} &= \frac{\vec w\tau}{3a}\left(1 - \frac{3}{5}\left(\frac{\vec w\tau}{3a}\right)^2\right)\;.
    \label{eq:m_nonlinear}
\end{split}
\end{equation}
This is just one of a long list of possible nonlinearities that may be considered. We choose to consider it above all others as it naturally controls the amplitude of the unstable modes, and without the addition of any further nonlinearities leads to stable long-term dynamics for this system. While $\vec m$ obeys equation (\ref{eq:m_nonlinear}), the velocity field $\vec w$ still follows equation (\ref{eq:general_eqmot_w}). This system cannot be solved analytically in general, so we will turn to a numerical method in the following section.

\subsection{Boundary Conditions}
Even though the nucleus usually looks like an ellipsoid, we reduce the complexity of the problem by modeling it as a sphere of radius $R$. 

We assume no-slip boundary conditions for both velocity fields $\vp,\;\vs$. In the case of the polymer, this is justified by the tethering of chromatin to the boudary by LINC complexes \cite{crisp2006coupling}, while no-slip boundary conditions for the solvent are standard for viscous fluids \cite{batchelor1967introduction}
\begin{equation}
    \vp_{\mathrm{tangential}} \big|_{r=R} = \vs_{\mathrm{tangential}} \big|_{r=R} = 0\;.
    \label{eq:noslip-condition}
\end{equation}
Turning now to normal components of velocity, we assume that no permeation of the boundary is possible by either polymer or solvent. We make this assumption despite the fact that the nuclear membrane is porous and lets some small molecules through \cite{alberts2017molecular}. We do so not only for simplicity, but also because over the seconds-timescale that we are interested in the volume of the nucleus is conserved \cite{chu2017origin}, thus the net flux of material through it is small. Finally, we treat the nuclear boundary as rigid and ignore its fluctuations, which are another source of potentially interesting effects \cite{chu2017origin}. All these simplifications result in the following boundary condition:
\begin{equation}
	\left[\phi\vec v^{\mathrm p} + (1-\phi)\vec v^{\mathrm s}\right]_{\mathrm{normal}}|_{r=R} = 0\;.
    \label{eq:noflow-condition}
\end{equation}
The meaning of this condition is simple: although neither component can go through the membrane, any one component, either polymer or solvent, can be approaching the membrane with some non-zero velocity provided that the other component at the same time departs from the wall, such that the exchange between them does not change volume.

\section{Results}

The spatial behavior of the linearized model (\ref{eq:w_m_transverse},\ref{eq:w_m_longitudinal}) is entirely contained in Laplacian operators. We will use the vector eigenfunction spectrum of the Laplacian to construct exact solutions for this model in the linear response regime. Given the restriction of the solutions to a closed bounded domain, the spectrum is discrete.

Once we turn to the nonlinear regime, it becomes impossible to solve the equations of motion exactly using these basis functions. However, we have constructed a numerical method for the full nonlinear model which exploits the basis we develop for the linearized version. This makes three-dimensional solutions of the full nonlinear PDE possible with minimal computational complexity. 

\subsection{Linearized model: an analytical solution}
\begin{figure}
    \centering
    \includegraphics[width=\linewidth]{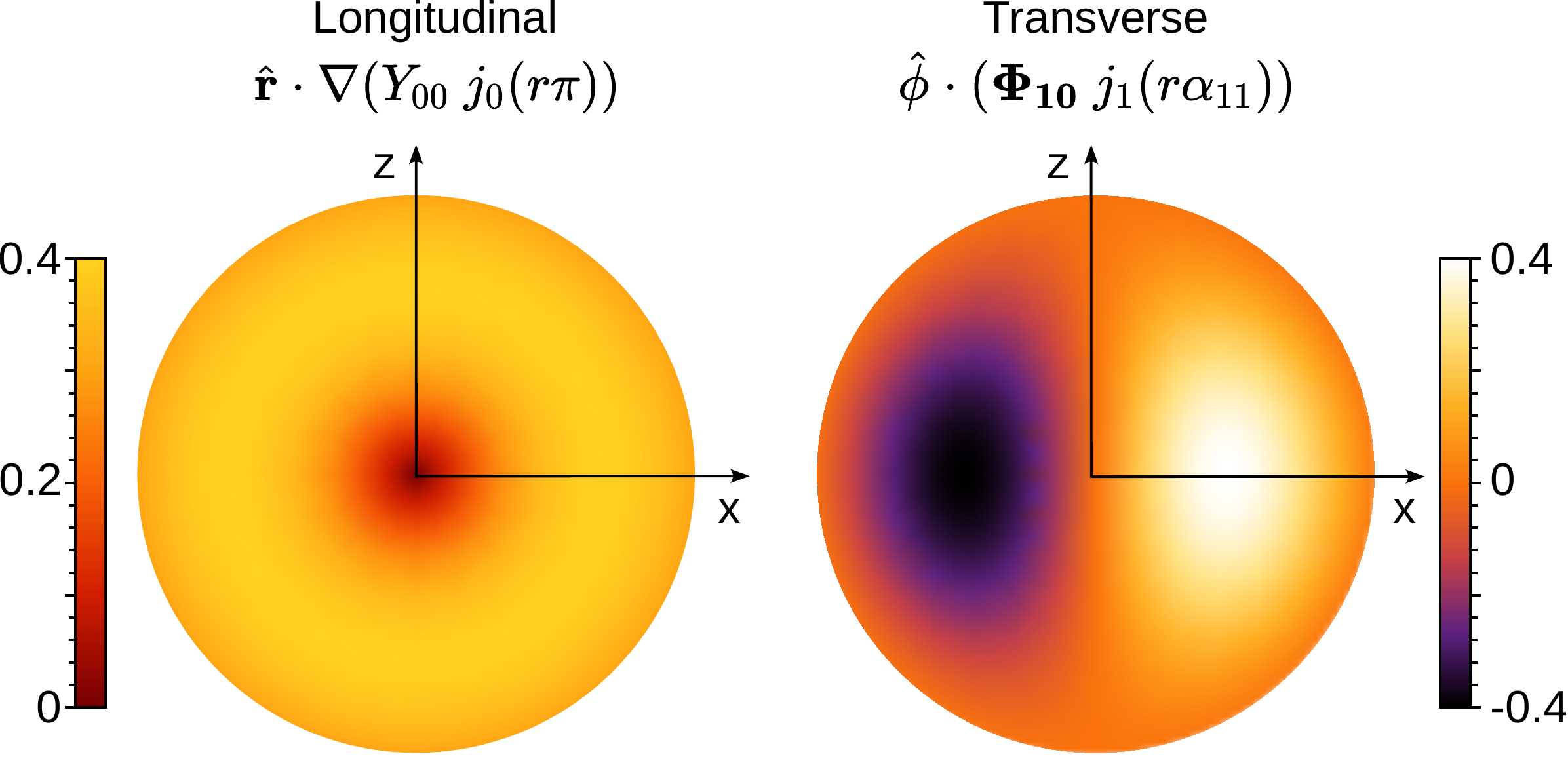}
    \caption{First excited longitudinal and transverse modes, shown along a vertical slice of the spherical domain. Since the vector dependence of these first modes is simple, we choose to plot only one component of the resulting vector fields, since the other components are simply $0$. In the case of the first longitudinal mode the flow is spherically symmetric, while the first transverse mode is axially symmetric about the $\hat{\vec z}$ axis.}
    \label{fig:modeplot}
\end{figure}
We begin by finding solutions in the linear response regime, where the distribution of motor orientations deviates weakly from uniform. This allows us to expand the flow fields in a basis which simplifies the equation of motion. If we consider curl-free ($\vec w_{\parallel}$) and divergence-free ($\vec w_{\perp}$) components of the velocity separately, then the dynamics of these modes separate and simplify, as shown in equations (\ref{eq:eqmot_split}).

The natural choice of basis functions to solve this system of equations needs to have two properties: they need to be eigenfunctions of the Laplace operator in spherical coordinates, to simplify the spatial dependence of the equations of motion, and they must be vectors to preserve the symmetry of the velocity fields. Such functions are already known, and are referred to as vector spherical harmonics (VSH) \cite{mcc1981morse}. They are constructed based on the well-known scalar spherical harmonics $Y_{lm}(\theta,\phi)$. There are several, closely related, possible definitions for VSH. We will use the following in this paper:
\begin{equation}
    \begin{split}
        \vec Y_{lm} &= Y_{lm} \hat{ \vec r},\\
        \vec \Psi_{lm} &= r\nabla Y_{lm},\\
        \vec \Phi_{lm} &= \vec r\times \nabla Y_{lm}\;.
    \end{split}
\end{equation}
From these, we construct curl-free and divergence-free components of $\vec w$. First, notice that for any scalar function $f(\vec r)$, we have $\nabla\cdot\left(f\vec \Phi_{lm}\right) =0$. Then recall that for a vector-valued function to be curl-free, it suffices that it be the gradient of a scalar. At every radial shell $r$, we expand the scalar function $f(\vec r)$ in spherical harmonics, with coefficients $f_{lm}(r)$. This gives us
\begin{equation}
\begin{split}
    \nabla f(\vec r) &= \sum_{lm}\nabla\left(f_{lm}(r)Y_{lm}\right) \\
    &= \sum_{lm} \left(\frac{\partial f_{lm}(r)}{\partial r} \vec Y_{lm} + \frac{f_{lm}(r)}{r}\vec \Psi_{lm}\right)\;.
\end{split}
\end{equation}
Therefore, we will solve equations (\ref{eq:eqmot_split}) using the following expansions:
\begin{equation}
\begin{split}
    \vec w_{\perp} &= \sum_{lm} a_{lm}(r,t)\vec \Phi_{lm};\\
    \vec w_{\parallel} &= \sum_{lm} \nabla\left(b_{lm}(r,t)Y_{lm}\right)\\
    &= \sum_{lm} \left(\frac{\partial b_{lm}(r,t)}{\partial r} \vec Y_{lm} + \frac{b_{lm}(r,t)}{r}\vec \Psi_{lm}\right)\;.
\end{split}
\label{eq:full_expansion}
\end{equation}
To identify what spatial dependence $a_{lm},b_{lm}$ must have, we search for eigenfunctions of the Laplacian which meet our boundary conditions. Here it is important to notice that by construction, the longitudinal velocity field $\vec w_{\parallel}$ automatically obeys condition (\ref{eq:noflow-condition}) due to the continuity of $\delta\phi$. Therefore, we only need to impose (\ref{eq:noslip-condition}) in the case of the longitudinal flows. As a result, we are searching for $a_{lm}(r,t),b_{lm}(r,t)$ which satisfy
\begin{equation}
\begin{split}
    &a_{lm}|_{r=R} = b_{lm}|_{r=R} = 0\\
    &\nabla^2\left(a_{lm}\vec\Phi_{lm}\right) = -k^2a_{lm}\vec \Phi_{lm}\;,\\
    &\nabla^2\left(b_{lm}Y_{lm}\right) = -k^2b_{lm}Y_{lm}\;,
\end{split}
\end{equation}
for some real number $k^2$. One such set of functions are spherical Bessel functions $j_l(x)$, which have the property $\nabla^2\left(j_l(kr)Y_{lm}\right) = -k^2j_l(kr)Y_{lm}$ in spherical coordinates. We choose $k$ such that the modes meet the boundary conditions at $r=R$. This results in the following expansion
\begin{equation}
\begin{split}
    \vec w_{\perp} &= \sum_{lmn} a_{lmn}(t)j_l\left(\frac{\alpha_{ln}r}{R}\right)\vec \Phi_{lm}\;,\\
    \vec w_{\parallel} &= \sum_{lmn} {b}_{lmn}(t)\nabla\left(j_l\left(\frac{\alpha_{ln}r}{R}\right)Y_{lm}\right)\;,\\
\end{split}
\label{eq:expansion_basis}
\end{equation}
where we have defined $\alpha_{ln}$ to be the $n$th zero of the $l$th order spherical Bessel function. Equipped with the basis (\ref{eq:expansion_basis}), it is now straightforward to insert the expansion into the equations of motion (\ref{eq:eqmot_split}), and solve for the time dependence of the coefficients $a_{lmn},b_{lmn}$:
\begin{subequations}
\begin{align}
    \label{eq:transverse_modes_stab}
    &\tau \dot a_{lmn}(t) = \frac{2\left(\epsilon - \lambda^2\left(\alpha_{ln}/R\right)^2\right)}{1+\lambda^2\left(\alpha_{ln}/R\right)^2}a_{lmn}(t)\;,\\
\begin{split}
    &\tau^2\ddot{b}_{lmn}(t) -2\frac{\left(\epsilon - (\lambda_d^2+\lambda_s^2)\left(\alpha_{ln} /R\right)^2\right)}{1+\lambda_s^2\left(\alpha_{ln}/R\right)^2}\tau\dot{b}_{lmn}(t)\\
    &\quad + 4\frac{\lambda_d^2\left(\alpha_{ln} /R\right)^2}{1+\lambda_s^2\left(\alpha_{ln}/R\right)^2}b_{lmn}(t) = 0\;.
    \label{eq:longitudinal_modes_stab}
\end{split}
\end{align}
\end{subequations}
The time dependent dynamics of these coefficients and the corresponding modes (\ref{eq:expansion_basis}) is very similar to what we have found previously for an unbounded domain, except of course the instability thresholds are significantly affected due the effect of boundaries. In the case of the transverse modes, the solutions of equation (\ref{eq:transverse_modes_stab}) are simple exponentials in time, with the sign of the coefficient on the right-hand-side of equation (\ref{eq:transverse_modes_stab}) determining stability. While at small $\epsilon$ the mode is stable, as soon as $\epsilon > \lambda^2\alpha_{ln}^2/R^2$ the mode becomes unstable. As expected, the finite-size effect is captured by the unitless ratio $\lambda/R$. If the domain is a lot larger than the mesh size, which corresponds to the thickness of the boundary layer needed to meet the no-slip condition, then the dynamics are similar to that of an infinite domain. However, if the domain is small, additional forcing is needed to excite these modes, since the viscous drag at the boundary will significantly dampen their motion. 

The time-dependence of the longitudinal modes, as in the infinite domain case (see \cite{eshghi2023alignment}), is that of a harmonic oscillator. The term in equation (\ref{eq:longitudinal_modes_stab}) which controls stability is the friction term, and it flips sign when $\epsilon$ is sufficiently large resulting in an apparent negative friction coefficient. As a result, the amplitude of the oscillations grow exponentially in time. The necessary forcing to drive this instability is far larger than the transverse case. Indeed, the critical value of $\epsilon$ in this case is $(\lambda_s^2 + \lambda_d^2)\alpha_{ln}^2/R^2$, and recall that $\lambda_s\gg\lambda$. Each of those terms represents a source of dissipation which must be overcome. $\lambda_s$ reflects the energy dissipation due to friction between polymer and solvent in these longitudinal modes. This term is larger than in the case of the transverse modes (in that case it was proportional to $\lambda$), because the transverse velocity of polymer is far smaller than that of solvent due to force balance and the fact that $\hp\gg\hs$. In the case of longitudinal flows however, the polymer can compress and move faster, leading to increased friction. Finally, $\lambda_d$ reflects the dissipation due to density change of the polymer.

It is worth considering what the first excited modes are in this geometry. The lowest-$\epsilon$ (easiest to excite) transverse mode is proportional to $\vec\Phi_{10} = \sin(\theta)\hat{\vec \phi}$. In this state, the chromatin is swirling around the $\hat z$ axis, with highest speed around the equator and decaying to $0$ at the poles. In contrast, the first longitudinal mode to be excited is proportional to $\vec Y_{00} = \frac{1}{4\pi}\hat{\vec r}$. This describes a radially-symmetric "breathing" motion of the chromatin, moving in and out of the center of the nucleus in an oscillatory manner. We show a slice of these first excited modes in Figure \ref{fig:modeplot}.

In summary, restoring physical parameters, the condition for the active force density is 
\begin{equation} f \rho > \frac{3 a \zeta T}{\gamma} + \frac{a \eta_s T\alpha_{11}^2}{\gamma R^2}  \label{eq:threshold_transverse}\end{equation}
for the transverse modes (the lowest allowable mode in this case is $l=1$, since $\vec \Phi_{00} = 0$). In the above, we have $\alpha_{11} \simeq 4.5$. For the longitudinal modes, the condition is 
\begin{equation}   f \rho  > \frac{3 a \zeta T}{\gamma}  + \frac{(1-\phi_0)^2 a\pi^2   }{R^2} \left[ \frac{3 T \eta^{\mathrm{p}}}{\gamma} + \frac{3}{2} K \phi_0 \right]\;,  \label{eq:threshold_longitudinal}
\end{equation}
where we have inserted the value of $\alpha_{01} = \pi$. These are the same expressions that we previously guessed \cite{eshghi2023alignment}, except the factors of $\alpha_{ln}^2$ were absent. We find that the addition of boundaries induces necessary gradients in the velocity field which the sources of activity must compete against. Furthermore, the necessary excess force is far larger if the motors are to compress and pump the polymer in a finite domain. Consistently with our previous results, these effects become less significant the larger the domain size is compared to the length scales $\lambda,\lambda_s,\lambda_d$.

\subsection{Nonlinear regime}

\begin{figure}
    \centering
    \includegraphics[width=0.8\linewidth]{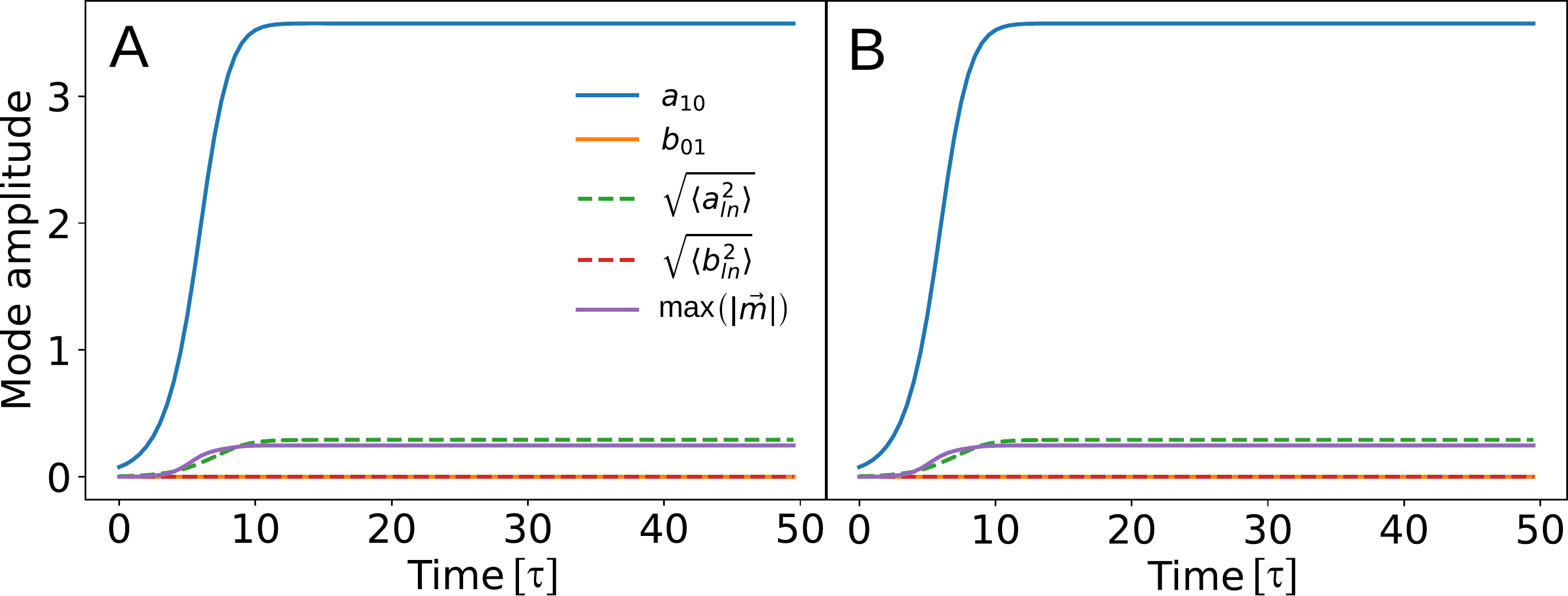}
    \caption{Transverse flows are stable in the nonlinear regime. A: Amplitude of the coefficients for the lowest transverse and longitudinal modes ($a_{10}$ and $b_{01}$ respectively), as a function of time when initialized in a pure state with $a_{10}(0)=0.1$. Nonlinearities mix multiple transverse modes together, but no longitudinal modes are excited. B: Amplitude of the same modes as in A, but with the initial conditions set to random numbers of magnitude $10^{-8}$, with $a_{10}(0) = 0.1$. The transverse steady-state is robust to this small perturbation.}
    \label{fig:stability_transverse}
\end{figure}
Now that we have identified the linear stability of this system with $\epsilon$ slightly above the critical point, we must consider the long-time behavior once the instabilities have grown and saturated the order parameter field $\vec m$. The nonlinear dynamics of this system, described by equations (\ref{eq:general_eqmot_w},\ref{eq:m_nonlinear}), have to be investigated using a numerical method. We developed such a method, which we describe in detail in the Appendix, section \ref{sec:numerical_solns}. To integrate the equations of motion, we convert the fields from a spatial representation to the basis (\ref{eq:expansion_basis}) at each time-step, then evolve the coefficients, before transforming back to spatial representation to evaluate the nonlinear time evolution for $\vec m$ shown above. We measured all length scales relative to the confinement scale $R$ and chose the numerical values $\lambda_d/R =10^{-1}$, $\lambda_s/R=10^{-2}$, and $\lambda/R=10^{-3}$.  This choice is consistent with our estimates that $\lambda_s \gg \lambda$ and also with numerical values of average mesh size in chromatin $\lambda \approx \unit[70]{nm}$  and $R \approx \unit[10]{\mu m}$. We set $\epsilon = 0.3$, above both thresholds of instability (\ref{eq:threshold_transverse},\ref{eq:threshold_longitudinal}).

The first question we set to explore is whether the modes identified in the linear response regime become stable once the nonlinearity becomes significant. We first test the transverse modes, by starting the system in a pure state $\vec w(t=0) = w_0j_1(r\alpha_{11}/R)\vec\Phi_{10}$, with $w_0 = 0.1$ and allowing it to evolve for $100$ time-steps of step size $\Delta t = 0.5$. Eventually the dynamics settle into a steady-state consisting of a superposition of transverse states, but no significant longitudinal excitations are observed, as shown in Figure \ref{fig:stability_transverse}A. This result is robust to small perturbations, which we test by initializing all other modes with uniformly distributed random values ranging between $\pm 10^{-8}$, as shown in Figure \ref{fig:stability_transverse}B.

\begin{figure}
    \centering
    \includegraphics[width=\linewidth]{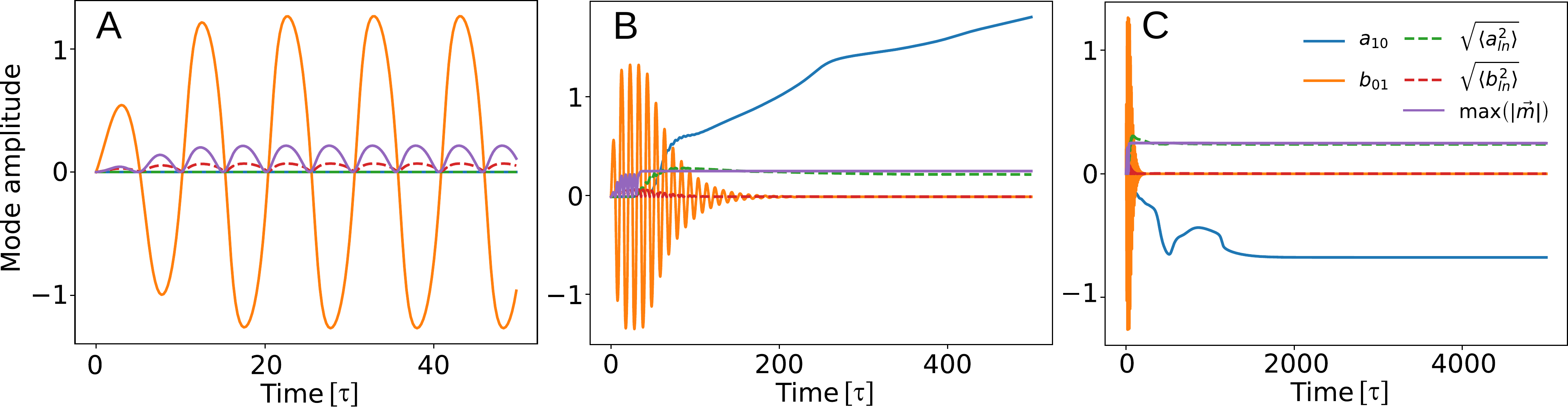}
    \caption{Longitudinal flows are unstable to perturbations in the nonlinear regime. A: Amplitude of the coefficients for the lowest transverse and longitudinal modes ($a_{10}$ and $b_{01}$ respectively), as a function of time when initialized in a pure state with $b_{10}(0)=0.1$. Nonlinearities mix longitudinal modes together, but the system remains free of transverse modes. B: Amplitude of the same modes as in A, but with the initial conditions set to random numbers of magnitude $10^{-8}$, with $b_{10}(0) = 0.1$. In this case, the transverse flows grow over time and eventually dominate the evolution of the system, suppressing all oscillations. C: Same as in B, but evolved over even longer timescales, to verify that the system does indeed reach a steady-state eventually.}
    \label{fig:stability_longitudinal}
\end{figure}

\begin{figure}
    \centering
    \includegraphics[width=0.5\linewidth]{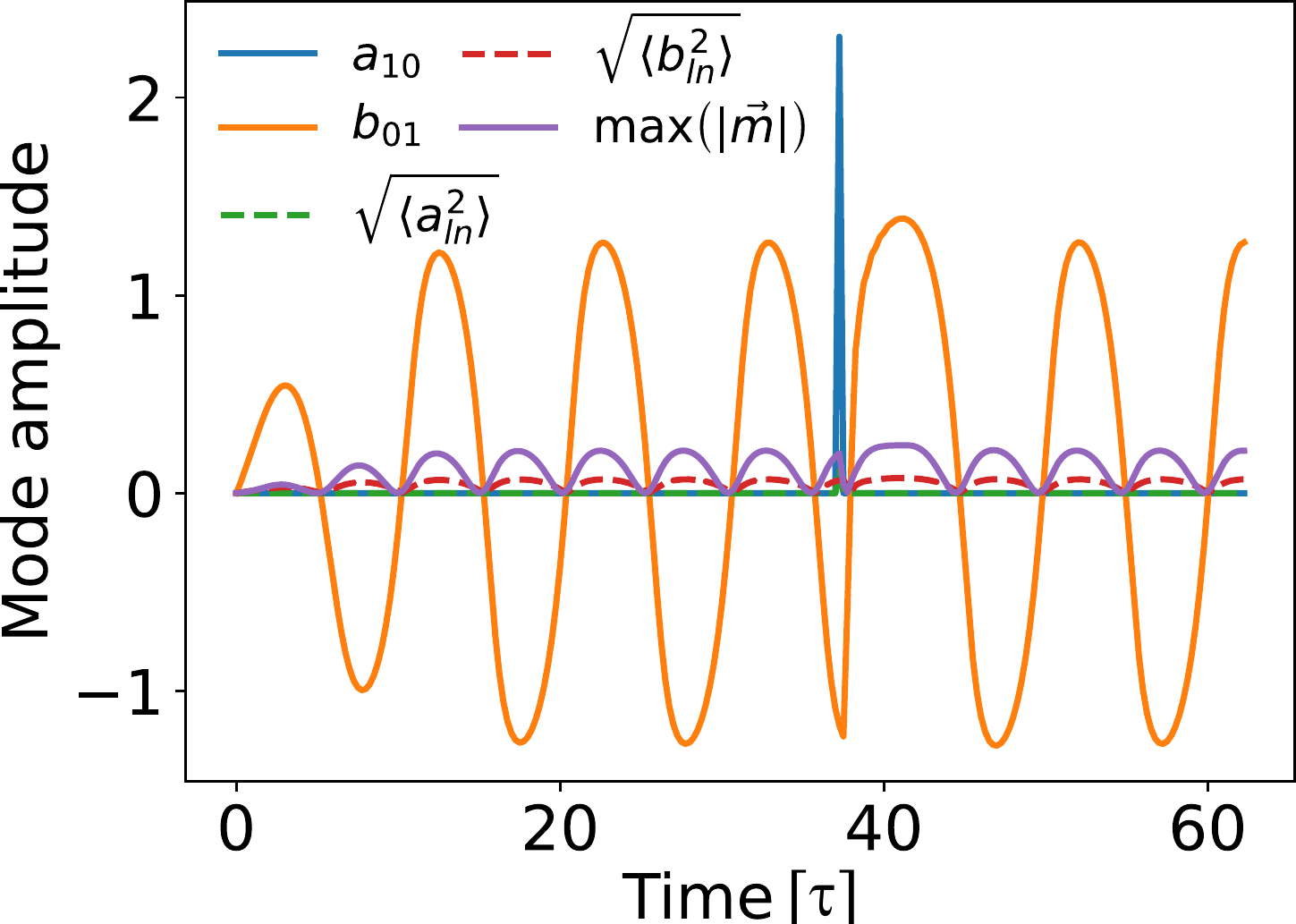}
    \caption{Longitudinal oscillations are stable to late perturbations. The system is prepared in the same manner as before, in a pure longitudinal state, but at a prescribed time the transverse mode $a_{10}$ is arbitrarily increased in magnitude to introduce a kick to teh system. For a short period the lowest longitudinal mode amplitude (orange) is affected, but it quickly settles back into its previous oscillatory behavior.}
    \label{fig:longitudinal_kick}
\end{figure}

We then perform the same analysis but with the longitudinal modes. We initialize the system in a pure state $\vec w(t=0) = w_0\nabla\left(j_0(r\alpha_{01}/R)Y_{00}\right)$ with $w_0 = 0.1$. To better resolve the oscillatory dynamics in this case we lower the time step to $\Delta t = 0.25\tau$, and we evolve for $200$ steps. The oscillations initially grow in magnitude until they settle at a maximum, at which point their amplitude remains steady for all times we integrated, as shown in Figure \ref{fig:stability_longitudinal}A. However, when we initialize the configuration with uniform random numbers ranging between $\pm 10^{-8}$ for all other modes, after a certain number of cycles the longitudinal oscillator gets taken over by the transverse steady-state modes, as shown in Figure \ref{fig:stability_longitudinal}B. To see where the evolution stabilizes we keep the system going for $2000$ time steps, and find that while the root-mean-squared magnitude of transverse modes reaches something close to a steady-state, individual modes such as the first coefficient $a_{10}(t)$ still evolve over those timescales, shown in Figure \ref{fig:stability_longitudinal}B. To verify that they eventually do reach a steady-state, we increase the time step to $\Delta t = 0.5\tau$ and evolve for $10^4$ steps, finding that the modes do eventually fully settle, albeit in a different steady-state each time due to the random initial conditions. One example evolution is shown in Figure \ref{fig:stability_longitudinal}C.
We test this stability further by allowing the longitudinal oscillations to fully develop before introducing transverse flows. We initialize the system in a pure longitudinal state as before, and allow it to evolve for $150$ time steps with step size $\Delta t = 0.25\tau$, which is equivalent to about three full cycles of the oscillator. Then, we introduce a transverse kick by setting the amplitude of the lowest transverse mode to a large value, $a_{10} = 3$. This briefly perturbs the oscillations as can be seen in the amplitude of $b_{01}$ right afterwards in Figure \ref{fig:longitudinal_kick}, but the system quickly settles back into its previously established resonance. Evidently, once the longitudinal oscillatory flows have settled in place, they are more robust to perturbations.

\section{Discussion \& Conclusions}

In this paper we have analyzed solutions of the active two-fluid equations of motion we derive in \cite{eshghi2023alignment}, but in a confined domain. As this model is intended to describe the active dynamics of the chromatin polymer and its solvent nucleoplasm, the addition of the confinement allows us to study the effect of the nuclear boundary on these dynamics. Specifically, we are interested in the kinds of active flows which may take place in such a restricted environment, and the ways in which the hierarchies of length scales (mesh size, screening length, osmotic relaxation given a characteristic time) may affect the stability of the dynamics. In our previous work \cite{eshghi2023alignment}, we performed estimates of the relevant length, time, and velocity scales using experimental values and found the theory to be consistent with experiments.

As estimated in that work, we find that the confinement length scale $R$ leads to factors of $1/R^2$ everywhere where the Laplace operator is present, resulting in confinement-dependent instability thresholds for both longitudinal and transverse modes. In the limit $R\rightarrow \infty$, we recover the unbounded results derived in \cite{eshghi2023alignment}. Furthermore, we find that the confinement size determines the size of the eigenmodes which get excited, which will be an interesting result to compare to experimental data on active chromatin flows.
We go beyond linear stability analysis by numerically integrating the equations of motion in a full three-dimensional domain. We find that the decomposition into transverse and longitudinal modes, which we performed for the sake of linear stability analysis, remains relevant in the nonlinear case. In particular we find that if initialized in a pure transverse or longitudinal state, the system will remain in that class of states (although the nonlinearity will mix transverse modes among one another and equivalently for longitudinal modes). Transverse modes are found to be stable under small perturbations, while longitudinal modes are stable under some but not others. It will be interesting, in a later study, to analyze in detail the basins of attraction and regimes of stability for all of these types of flows in the nonlinear regime. In sum, we find that the minimal active two-fluid model we have developed, exhibits rich and nontrivial behavior.

It is worth asking whether the choice of no-slip boundary conditions is too restrictive to be an accurate description of the physics of chromatin in the nucleus. After all, the nuclear envelope has a complex nature and is itself actively fluctuating \cite{chu2017origin}, which will significantly affect the possible motions of the fluids within. Furthermore, in this model we are not allowing any change in the total nuclear volume, or permeation of solvent through the boundary, both of which occur in real biological systems. These, and many other improvements on the model, could be added in a systematic way to this hydrodynamic description of chromatin dynamics, hopefully increasing our understanding of the complex physics of the genome.
\backmatter 

\bmhead{Acknowledgements}
AZ is grateful for support from the NSF Grants CAREER PHY-1554880, CMMI-1762506, PHY-2210541, DMS-2153432, and NYU MRSEC DMR-1420073. This research was supported in part by the National Science Foundation under Grant No. NSF PHY-1748958.
AZ and AYG acknowledge useful discussions with participants of the 2020 virtual KITP program on "Biological Physics of Chromosomes". Numerical solutions were calculated using New York University (NYU) High Performance Computing cluster.

\bmhead{Data Availability}
Data sharing not applicable to this article as no datasets were generated or analysed during the current study.

\begin{appendices}

\section{Completeness of solenoidal solutions}\label{sec:completeness}In the main text, we simply stated that the solenoidal solutions, that is those solutions with $\nabla\cdot\vec w = 0$, are spanned by the expansion
\begin{equation}
    \vec w_{\perp} = \sum_{lmn} a_{lmn}j_{l}\left(\frac{\alpha_{ln}r}{R}\right)\vec \Phi_{lm}\;.
    \label{eq:solenoidal_phi_expansion_si}
\end{equation}
However, we did not show that there is no combination coming from the basis functions $\vec \Psi_{lm},\; \vec Y_{lm}$. Here we will demonstrate that this is in fact the case.

Consider one term in the expansion $\vec w_{\perp} =  E^r(r)\vec Y_{lm} + E^{\theta}(r)\vec \Psi_{lm}$. The divergence of this field is
\begin{equation}
\begin{split}
    \nabla\cdot \vec w_{\perp} &= \left(\partial_r E^r + \frac{2}{r}E^r - \frac{l(l+1)}{r}E^{\theta}\right)Y_{lm}\\
    &= 0\;.
\end{split}
\end{equation}
Since the basis functions must be eigenfunctions of the spherical Laplacian and meet the no-slip boundary condition at $r=R$, we must have
\begin{equation}
    E^r = C j_l\left(\frac{r\alpha_{ln}}{R}\right)
\end{equation}
for some constant C. Enforcing the divergence-free condition, we solve for $E^{\theta}$:
\begin{equation}
    E^{\theta} = C\left(\frac{2+l}{r}j_l\left(\frac{\alpha_{ln}r}{R}\right) - \frac{\alpha_{ln}}{R}j_{l+1}\left(\frac{\alpha_{ln}r}{R}\right)\right)
\end{equation}
which cannot be made to meet the boundary condition at $r=R$. Thus we have shown that the expansion (\ref{eq:solenoidal_phi_expansion_si}) spans possible divergence-free functions that meet our boundary conditions and are eigenfunctions of the spherical Laplace operator.

\section{Numerical methods}
\label{sec:numerical_solns}
We wrote a numerical scheme in \verb|python| which successfully integrates the equations of motion
\begin{equation}
\begin{split}
	&\left(1-\lambda_s^2\nabla\nabla\cdot + \lambda^2\nabla\times\nabla\times\right)\tau\partial_t \vec w - 2\lambda_d^2\nabla\nabla\cdot \vec w\\
	&\quad\quad= \frac{f\rho}{\zeta}\tau\partial_t\vec m\\
	&\tau\partial_t \vec m = 2(\vec m_{eq}(\vec w)- \vec m)\\
	&\vec m_{eq} = \frac{\vec w\tau}{3a}\left(1 - \frac{3}{5}\left(\frac{\vec w\tau}{3a}\right)^2\right)\;.
\end{split}
\end{equation}

Given the initial fields $w$ and $m$ at time $t=0$, we first compute their decomposition into normal modes, i.e., eigenfunctions (\ref{eq:expansion_basis}). This process is outlined in further detail below. These modes are uncoupled in the equation of motion (\ref{eq:general_eqmot_w}) for $\vec w$, and so we can evolve the normal modes for $\vec w$ one time-step directly using equations (\ref{eq:transverse_modes_stab},\ref{eq:longitudinal_modes_stab}), discretized using an Adams-Bashforth two-step scheme. The resulting difference equations for $a_{nlm},b_{nlm}$ are shown below. 

However, at each time-step, the field $\vec m$ must also be updated, following equation (\ref{eq:m_nonlinear}). To do this, the normal modes are reassembled into the full spatial dependence of $\vec m,\vec w$, and equation (\ref{eq:m_nonlinear}) is evolved directly. Then the procedure above is repeated.

Throughout this scheme, we used the \verb|scipy.special| package for the definition of the spherical Bessel functions, but the vector spherical harmonics were implemented through the freely available package \verb|shtns|, detailed in the work \cite{schaeffer2013efficient}.

\subsection{Time evolution}
The time evolution was implemented using an Adams-Bashforth two-step scheme for the source terms, combined with a mixed Crank-Nicolson scheme in the case of the equation for $b_{nlm}$ to ensure stability (for the details on the definitions of these schemes, see \cite{press1988numerical}):
\begin{equation}
\begin{split}
    &a_{nlm}^{t+1} = \Delta t\frac{\epsilon+1}{1+\left(\frac{\lambda\alpha_{ln}}{R}\right)^2}\left(\frac{3}{2}\left(p_{\perp nlm}\right)^t - \frac{1}{2}\left(p_{\perp nlm}\right)^{t-1}\right)\\
    &b_{nlm}^{t+1} = \Delta t\frac{B_{ln}}{1+\Delta t A_{ln}}\left(\frac{3}{2}\left(p_{\parallel nlm}\right)^t - \frac{1}{2}\left(p_{\parallel nlm}\right)^{t-1}\right) \\
    & \quad \quad +\frac{1-\Delta t A_{ln}}{1+\Delta t A_{ln}}b_{nlm}^{t-1}\\
    &A_{ln} = \frac{\left(\lambda_d\alpha_{ln}/R\right)^2}{1+\left(\lambda_s\alpha_{ln}/R\right)^2},\; B_{ln} = \frac{\epsilon+1}{1+\left(\lambda_s\alpha_{ln}/R\right)^2}\;,
\end{split}
\end{equation}
where we have defined the field $\vec p = \partial_t \vec m$ as shorthand.

\subsection{Mode decomposition}

The remaining task is the conversion from spatial to spherical harmonic representations, and vice versa. The angular dependence of the decomposition is taken care of by the routines within the \verb|shtns| package, however we had to implement our own decomposition of the radial part into Bessel functions.

We will illustrate this whole process with a simple example. Consider a vector field $\vec f(\vec r)$, which we seek to decompose into the spherical harmonic and Bessel function basis, giving coefficients $f_{\perp nlm}, f_{\parallel nlm}$. We perform this decomposition in two steps. First, we loop over radial slices $\{r_i\}$, and for each of these values we call \verb|shtns| to decompose the vector field into spherical components
\begin{equation}
\begin{split}
	\vec f(r_i,\theta,\phi) &= \sum_{lm} Q_{lm}(r_i)\vec Y_{lm}(\theta,\phi)\\
	&+ S_{lm}(r_i) \vec \Psi_{lm}(\theta,\phi)\\
	& + T_{lm}(r_i)\vec \Phi_{lm}(r_i)\;,
\end{split}
\end{equation}
which gives us a discrete radial representation of the functions $Q_{lm}(r),S_{lm}(r),T_{lm}(r)$. We can directly read off that $T_{lm}(r) = \sum_{nlm}f_{\perp nlm}j_l(\alpha_{ln}r/R)$. However, the definition of the longitudinal modes is more ambiguous. This is resolved by considering the gradient of a generic scalar field $\rho(\vec r)$ in terms of VSH:
\begin{equation}
    \nabla\left(\rho_{lm}(r)Y_{lm}\right) = \frac{\rho_{lm}(r)}{r}\vec \Psi_{lm} + \rho'_{lm}(r)\vec Y_{lm}
\end{equation}
Therefore, for a given spherical harmonic mode we must have
\begin{equation}
    rS_{lm}(r) = \int_{0}^{r}Q_{lm}(r')d r' = \sum_{nlm}f_{\parallel nlm}j_{l}(\alpha_{ln}r/R)\;.
\end{equation}
For the mode where $l=0,m=0$, we decompose $Q_{00}$ to obtain the coefficients of the \textit{derivative} of $j_0(r\alpha_{0n}/R)$, since $\Psi_{00} = 0$. Otherwise we decompose the function $rS_{lm}(r)$ to get coefficients of $j_l(r\alpha_{ln})$, as this is computationally simpler. Both in the case of longitudinal and transverse flows, the task is always reduced to the decomposition of a function defined over the interval $[0,1]$ into spherical Bessel functions
\begin{equation}
\begin{split}
    f_l(x) &= \sum_{n} f_{ln}j_l\left(x\alpha_{ln}\right), \; l>0\\
    f_0(x) &= \sum_n f_{0n}j_0'\left(n\pi x\right) = -\sum_n n\pi f_{0n}j_1(n\pi x)\;.
\end{split}
\label{eq:radial_decomp_abstract}
\end{equation}
where we have suppressed the index $m$ as it does not affect this part of the calculation, and inserted the known roots $\alpha_{0n} = n\pi$. To find the components $f_{ln}$, we exploit the following orthogonality relation
\begin{equation}
\begin{split}
    &\int_0^1 j_l(x\alpha_{ln})j_l(x\alpha_{lm})x^2 d x = \\
    &\quad \frac{\delta_{nm}}{2}\times
    \begin{cases}
        -j_{l-1}(\alpha_{ln})j_{l+1}(\alpha_{ln}),\; l>0\;.\\
        \frac{1}{(n\pi)^2},\;l=0
    \end{cases}
\end{split}
\end{equation}
Multiplying equations \ref{eq:radial_decomp_abstract} by $x^2$ and integrating, we obtain
\begin{equation}
\begin{split}
    f_{ln} &= -2\frac{\int_0^1 x^2 j_l(x\alpha_{ln})f_l(x)d x}{j_{l-1}(\alpha_{ln})j_{l+1}(\alpha_{ln})},\;l>0\\
    f_{0n} &= -2n\pi\int_{0}^1x^2j_1(n\pi x)f_0(x)d x\;.
\end{split}    
\end{equation}
These integrals are then implemented in our code. To perform the integrals over the discrete set of radii $\{r_i\}$, we used Simpson's rule as implemented in \verb|scipy.integrate|. 

The inverse operation, going from coefficients to the full spatial dependence, is far simpler: for each $l$, we sum over Bessel functions multiplied by their coefficients $f_{ln}$. This gives us the radial dependence of the coefficients $Q_{lm}(r),S_{lm}(r),T_{lm}(r)$. We then use the \verb|shtns| package to convert those back to spatial representations, radial shell by radial shell.
\end{appendices}
\bibliography{sources}
\end{document}